# Long-Term Care Facilities as a Risk Factor for Death Due to COVID-19

Neil Gandal, Matan Yonas, Michal Feldman, Ady Pauzner, and Avraham Tabbach[1]

June 23, 2020

## Abstract

A large percentage of the deaths from COVID-19 occur among residents of long-term care facilities. There are two possible reasons for this phenomenon. First, the structural features of such settings may lead to death. Alternatively, it is possible that individuals in these facilities are in poorer health than those living elsewhere, and that these individuals would have died even if they had not been in these facilities.

Our findings show that, controlling for the population density, the percentage of older adults in the population, and the number of hospital beds per capita, there is a significant positive association between the number of long-term care beds (LTCB) per capita and COVID-19 mortality rates.

This finding provides support for the claim that long-term care living arrangements (of older people) are a significant risk factor for dying from COVID-19.

[1] Neil Gandal, Tel Aviv University, School of Economics, and CEPR (gandal@tauex.tau.ac.il)
Matan Yonas, Tel Aviv University, School of Economics, (matany@mail.tau.ac.il)
Michal Feldman, Tel Aviv University, School of Computer Science, (michal.feldman@cs.tau.ac.il)
Ady Pauzner, Tel Aviv University, School of Economics, (pauzner@tauex.tau.ac.il)
Avraham Tabbach, Tel Aviv University, Law Faculty, (adtabbac@tauex.tau.ac.il)
We are grateful to Liat Ayalon, Noam Gandal, Mark Last, Analia Schlosser, Sarit Weisburd, Charles Wyplosz, Yishay Yaffe, and seminar participants at Tel Aviv University for helpful suggestions and comments. This work was partially supported by Len Blavatnik and the Blavatnik Family foundation.

1. Introduction

A large percentage of the deaths worldwide from COVID-19 have occurred among residents of long-term care institutions.[2] *Euronews* reported that deaths due to COVID-19 among such long-term care residents could account for more than 50% of all COVID-19 deaths in Europe.[3] According to an article in *The Guardian*, data from the Kaiser Family Foundation indicates that COVID-19 deaths among long-term care residents account for more than 50% of all deaths attributed to COVID-19 in fourteen states in the United States. Additionally, the same article notes that in the state of New Hampshire, 72% of COVID-19 deaths occurred among those living in long-term care settings.[4] Overall, according to the *New York Times*,[5] more than one third of the deaths in the United States from COVID-19 have occurred among long-term care residents. The U.S. Center for Disease Control and Prevention (CDC) has formally stated that generally, people 65 years and older, and in particular "People who live in a nursing home or long-term care facility" are at high-risk for severe illness from COVID-19.[6]

There are two possible explanations for the higher COVID-19 mortality rates in long-term care facilities:

1. The structural features of such settings, such as a communal living area, multiple residents in a room, care provided by multiple caregivers to multiple care recipients, etc., can lead to a greater number of deaths.
2. Individuals in these facilities are in poorer health than those outside of such facilities and they would have been likely to die even had they had not been in these facilities.

Each of these has different policy implications.

This paper examines the two competing explanations by studying the association between long-term care beds per capita in a country and COVID-19 deaths per capita. Using country-level data from Europe,[7] and controlling for the percentage of older adults in the population, the population density and the number of hospital beds per capita, we find that there is a significant positive association between the number of long-term-care beds per capita[8] and COVID-19 mortality rates in European countries. This finding supports the thesis that living in long-term care facilities presents a significant mortality risk factor for older people contracting COVID-19.

---

[2] See Comas-Herrera et. al. (2020).
[3] https://www.euronews.com/2020/04/17/coronavirus-care-homes-could-be-where-over-half-of-europe-s-covid-19COVID-19-deaths-occur-says-new. See also the report by Comas-Herrera et. al., 2020.
[4] https://www.theguardian.com/us-news/2020/may/11/nursing-homes-us-data-coronavirus
[5] https://www.nytimes.com/2020/05/12/business/nursing-homes-coronavirus.html
[6] https://www.cdc.gov/coronavirus/2019-ncov/need-extra-precautions/groups-at-higher-risk.html
[7] All data sources can be found in the Appendix.
[8] We do not have data on how many people aged seventy-five and older are living in long-term care settings. Therefore, we use long-term care beds per capita as a proxy for older persons living in long-term care facilities. Since such facilities are typically "full to capacity," we believe is an excellent proxy for the number of people living in such setting.

Our results also provide a partial, preliminary explanation as to why the death rates from COVID-19 differ so widely among the European Countries. This issue needs to be explored in more depth once there is more detailed data available, including additional countries, regional level data, and more. Therefore, there is a need for future research on this question.

2.  Analysis Using European Countries

This research seeks to examine the factors that are associated with deaths per capita from COVID-19, and, in particular, long-term care beds per capita. Before we write down and estimate an econometric model, we examine the raw data.

Data

The data employed in the study are

- Deaths_cap = deaths from COVID-19 per million residents through mid-May 2020
- LTCB_cap = number of long-term care beds per million residents[9]
- Per_75 = the percentage of the population aged 75 and older[10]
- Pop_den = the population density: residents per square kilometer.
- Hosp_cap = Hospital Beds per million residents

We included all European countries for which we have data on long-term beds per capita. We have data on thirty-two of the thirty-six European countries with more than 600,000 residents.[11] Figure 1 shows a graph of deaths per capita in relation to long-term care beds per capita for these countries. The figure shows that there is a large difference in the number of long-term care beds per capita. The figure also suggests that there is a positive association between Long-term care beds per capita and COVID-19 Deaths per capita.[12]

---

[9] The European Health Information Gateway, which is part of the European Regional Office of the World Health Organization is the source for the data used in this study and for the definition of long-term care beds. Their definition for long-term care beds is "beds available for people requiring long-term care in institutions (other than hospitals.) The predominant service component is long-term care and the services are provided for people with moderate to severe functional restrictions." The quote is from the European Health Information Gateway, https://gateway.euro.who.int/en/indicators/hfa_491-5101-number-of-nursing-and-elderly-home-beds/, which served as the data source for all countries except Portugal for this study. Most of the data on beds from our data source is from 2015, the most recent year available. See the Appendix for details. See the source for the detailed explanation. Although the quality of the settings, and their structures may differ, the nature of the facilities included is well-defined by the agency.

[10] All of the results on LTCB per capita obtain if we use "65 and older" or "70 and older" instead of "75 and older." The "fit" of the regressions is best when we employ "75 and older." Hence, we use this variable.

[11] Two small island countries, Iceland and Malta, were excluded from the analysis. We also excluded Russia and Turkey since these countries are primarily in Asia. Our results are qualitatively unchanged if we include Russia and Turkey.

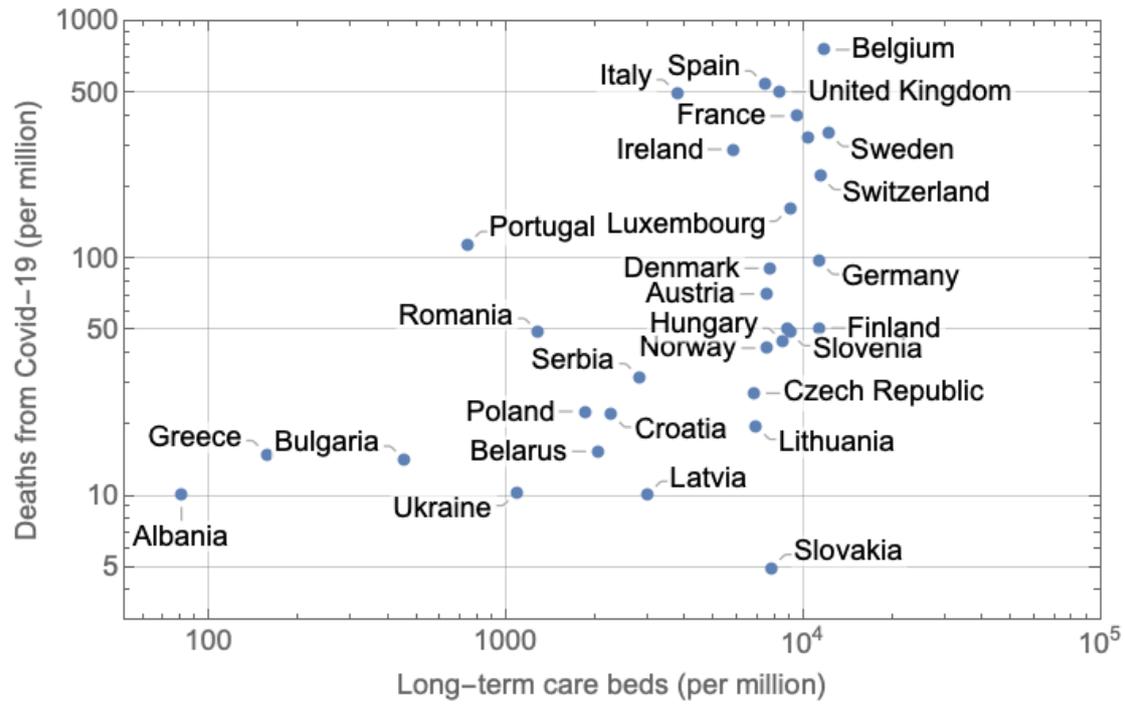

Figure 1: COVID-19 Deaths per capita in relation to LTCB per capita: 32 Countries: logarithmic scale.

Descriptive Statistics are shown in Table 1.

Table 1: Descriptive Statistics (N=32)

| Variable | Mean | Std. Error | Minimum | Maximum |
| --- | --- | --- | --- | --- |
| Deaths per capita | 153.3 | 197.9 | 5.0 | 760.0 |
| LTCB per capita | 6,196.3 | 3,945.5 | 80.7 | 12,140.0 |
| Per_75 | 0.088 | 0.017 | 0.06 | 0.12 |
| Population Density | 123.7 | 96.6 | 14.5 | 412.5 |
| Hospital Beds per capita | 5,896.8 | 1968.4 | 2200.0 | 11,200.0 |

In Table 2 below, we show the correlations among the above variables. The Table shows that long-term beds per capita is positively correlated with COVID-19 deaths per capita (0.45.) The table also shows that deaths per capita are positively correlated with the percentage of the population aged 75 and over (correlation 0.24, ) and with the population density (correlation

---

[12] While not all the country names not appear in Figure 1, all countries are included in the figure.

0.60). The table also shows that deaths per capita are negatively correlated with hospital beds per capita (correlation -0.39.)

Table 2: Correlations among the Variables (N=32)

|  | Deaths per cap | LTCB per cap | Per_75 | Hosp Beds per cap |
|---|---|---|---|---|
| Deaths per capita | 1.00 |  |  |  |
| LTCB per capita | 0.45 | 1.00 |  |  |
| Per_75 | 0.24 | 0.17 | 1.00 |  |
| Population Density | 0.60 | 0.38 | 0.03 | 1.00 |
| Hospital Beds per capita | -0.39 | -0.04 | -0.15 | -0.13 |

We now turn to the Econometric Analysis.

Econometric Analysis

The first structural equation has COVID-19 Deaths per capita (Deaths_cap) on the left-hand side. On the right side we include long-term care beds per capita (LTCB_capita,) the percentage of the population aged 75 and older (Per_75), and hospital beds per capita (Hosp_cap,) all of which are exogenous.[13] We also include COVID-19 cases per capita (Cases_cap) in the equation. The first equation is thus

1. $Deaths\_cap = \beta_0 + \beta_1 \ast LTCB\_cap + \beta_2 \ast Per\_75 + \beta_3 \ast Hosp\_cap + \beta_4 \ast Cases\_cap + \varepsilon$,
where $\varepsilon$ is the error term.

Cases per capita is endogenous and likely depends on the two exogenous variables in Equation 1, as well as on population density (pop_den), which is also exogenous, and on tests performed for COVID-19 per capita (Tests_cap,) which itself may be endogenous. The second structural equation is thus

2. $Cases\_cap = \alpha_0 + \alpha_1 \ast LTCB\_cap + \alpha_2 \ast Per\_75 + \alpha_3 \ast pop\_den + \alpha_4 \ast Tests\_cap + \xi$,
where $\xi$ is the error term.

Tests per capita depends on exogenous variables like institutional features of the country, and government policy, which we can reasonably assume are uncorrelated with the other

---

[13] We would have liked to include the percent of the population with two or more preexisting conditions in the older people in the population as an explanatory variable, but we do not have data on this variable. It seems reasonable to assume that this variable is uncorrelated with the exogenous variables used in our analysis.

exogenous variables and can be viewed as part of the error term. But tests per capita is also likely a function of Cases per capita, that is, more cases lead to more tests. Thus, the third structural equation is

3. Tests_cap = $\gamma_0$ + $\gamma_1$*Cases_cap + $\mu$, where $\mu$ is the error term.

We are interested in the association between long-term care beds per capita (LTCB_cap) and deaths per capita (deaths_cap.) Since the number of long-term care beds per million residents is exogenous, we can solve for the reduced form of the above three-equation structural model. We can then estimate the relevant "reduced form" equation with Deaths per capita as the left-hand side variable:[14]

4. Deaths_cap = $\varphi_0$ + $\varphi_1$*LTCB_cap + $\varphi_2$*Per_75 + $\varphi_3$*pop_den + $\varphi_4$*hosp_cap + $\omega$

Hence, we can estimate equation (4) using ordinary squares, and the estimates will be unbiased.

From Figure 1, it appears that the variance of the dependent variable increases with the number of long-term care beds. Hence, we use robust standard errors, which is the way to address heteroscedasticity.

In Column 1 of table 1, we estimate a linear model using equation (4.) In this case, we find that the estimated coefficients on long-term beds per capita and population density are positive and statistically significant at the 95 percent level of confidence. The estimated coefficient on the percent of the population 75 and older is positive, but not statistically significant. The coefficient on Hospital Beds per capita is negative and statistically significant at the 95 percent level of confidence.

In column 2 of Table 3, we estimate (4) using a log/log model. The regression uses the natural logarithm of deaths per million residents from COVID-19 as the dependent variable and the natural logarithm of LTCB per capita, the natural logarithm of population density, the natural logarithm of the percent of the population 75 and older, and the natural logarithm of hospital beds per capita as explanatory variables.

---

[14] There are also reduced form regressions for Cases per capita and Tests per capita as a function of the same exogenous variables in equation 4. But we are interested in the association between long-term care beds per capita and death per capita.

Similar to the specification in Column 1 of Table 3, the results in column 2 show that the estimated coefficients for both long-term beds per capita and population density are positive and both are statistically significant, in this case at the 99 percent level of confidence. The coefficient on the percent of the population seventy-five years and over is positive, but not significant, while the coefficient on hospital beds per capita is negative and statistically significant at the 99 percent level of confidence. Thus the results are qualitatively similar to those using the linear model in Column 1 of the table.

The table also shows that the four factors in the log/log model explain 68 percent of the variation in the death rate for European countries (versus 54 percent for the linear model.) This is a very high percentage for such a small number of factors. Hence, we believe that the fit of this model is quite good. While both the linear model and the log/log model provide qualitatively similar results, the log/log model is our preferred specification since it provides the better fit and makes sense intuitively.[15]

The coefficient on the natural logarithm of LTCB per capita is an elasticity. Thus, the 0.58 coefficient from the second regression in Table 3 means that a one percent increase in the number of long term care beds per capita is associated with a 0.58 percent increase in deaths per capita.

Robustness Analysis

LTCB could be a function of wealth, such that wealthier countries will have more beds. If this were driving our results this would imply that GDP per capita is positively correlated with deaths per capita. While, it seems clear that increases in GDP per capita will not directly lead to an increase in the death rate, our story suggests that it is not the earnings themselves but the behavior of wealthier countries (specifically in terms of their treatment of older people) that drives our outcomes.

- In any case, when we add GDP per capita as an explanatory variable, the results are qualitatively unchanged as shown in Column 4 in Table (3.) In particular, the estimated coefficient on long-term care beds per capita is positive and statistically significant at the 99% level of confidence using the preferred Log/Log model. The estimated coefficient on GDP per capita is positive, but not statistically significant.

---

[15] Other things being equal, the marginal long-term care "resident" in a country with more LTCB per capita is likely healthier than the marginal long-term care resident in a country with fewer LTCB per capita, and thus less likely to die if infected. This suggests that the relationship between LTCB per capita and deaths would be such that the death rate would be increasing at a decreasing rate as the number of LTCB per capita increases. In such a case, a log/log model is ideal.

- GDP per capita might lead to more deaths if wealthier countries find it more difficult to social distance due to economic activity, which in turn might lead to more deaths. The ideal way to address this is to include a variable on "mobility" in the regression.

  Our mobility index was calculated by taking the average of the following "Google mobility" indices (I) "Residential", (II) "Workplace", (III) "Transit", and (IV) "Retail & Recreation".[16] We do this, and report the results in column 4 of table 3.[17] We also include GDP per capita in this regression. We find that our results are again qualitatively unchanged. In particular, the estimated coefficient on LTCB beds per capita is positive and statistically significant at the 97 percent level of confidence. The estimated coefficients on mobility and GDP per capita are positive, but not statistically significant.

- Thus, even after controlling for GDP per capita, and mobility, as well as the other factors, LTCB per capita is positively associated with death per capita and the effect is statistically significant. Hence, we are confident that our results are robust.

- In summary, controlling for other factors that might affect the death rate, the estimated coefficient on long-term care beds per capita is positive and statistically significant at the 95 percent level or a higher level. Thus our preliminary results suggest that long-term care beds per capita emerges as a risk factor for death from COVID-19.

---

[16] See the appendix for the source of the mobility data.

[17] We do not have data for Albania so there are 31 observation in the regression in column 4. The correlation between mobility and LTCB is virtually zero (0.03.) Other indices of mobility give similar results.

Table 3: Estimates from Equation (4)[18]

|  | Linear Model (1) | Log/Log (2) | Log/Log (3) | Log/Log (4) |
|---|---|---|---|---|
| LTCB_capita | 0.012** (0.0055) | 0.58***(0.086) | 0.39***(0.14) | 0.40**(0.17) |
| Per_75 | 1716.0 (1543.6) | 1.36 (1.12) | 1.35 (1.02) | 1.18 (1.20) |
| Pop_den | 0.97** (0.37) | 0.67***(0.11) | 0.63*** (0.13) | 0.57***(0.17) |
| Hosp_cap | -0.030** (0.014) | -1.76***(0.33) | -1.51***(0.37) | -1.69***(0.38) |
| GDP_cap |  |  | 0.64 (0.41) | 0.55 (0.40) |
| Mobility |  |  |  | 0.39 (0.81) |
| $R^2$ | 0.54 | 0.68 | 0.71 | 0.70 |
| N | 32 | 32 | 32 | 31 |

3. The Regions in Italy

For research purposes, having data on long-term care beds per capita for regions within a country, and not just for the country as a whole, is ideal, as there is more similarity in other (unobserved) aspects within a country than among countries. The one country that did have such regional data available was Italy. This information also existed for us to break down the data for Italy into the smaller sub-regions, which we denote as "counties." The correlation of deaths per capita and long-term care beds per capita is 0.70 when all counties are included. Since the northern part of Italy had many more deaths per capita, the correlation between COVID-19 deaths and living in long-term care facilities was again calculated excluding the northern counties. The result of 0.74 was very close to that of the country as a whole. This suggests that even when excluding the northern regions of Italy from consideration, there is strong positive relationship between mortality rates per capita and long-term care beds per capita. See Figure 2, which shows the data for Italian counties.

---

[18] ROBUST Standard Errors in parentheses: ***= significant at 99% level, **= significant at the 95% level, and *= significant at the 90% level.

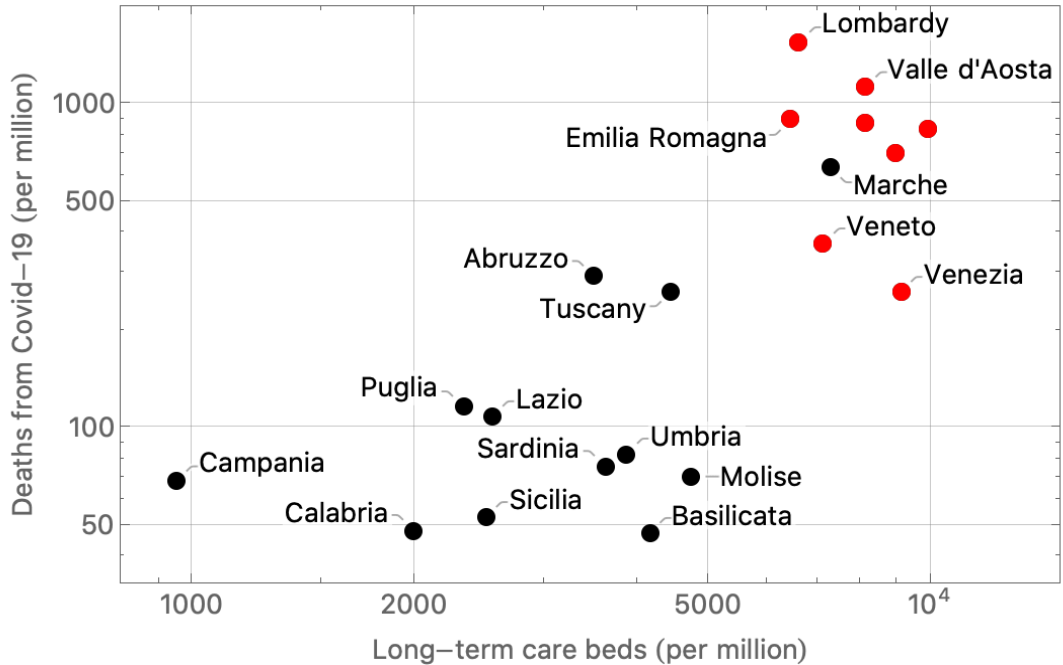

Figure 2: COVID-19 Deaths per capita in relation to long-term care beds per capita, Italian counties, northern counties in "red" (logarithmic scale.)

4. Discussion and Further Work

Controlling for the proportion of adults "aged 75 and older," the population density, and hospital beds per capita, the number of long-term care beds per capita is positive and statistically significant in explaining the differences in the deaths per capita for countries in Europe. This suggests that the structural features of such settings are associated with death from COVID-19. In European countries with more long-term care beds per capita, the death rate from COVID-19 is higher. These findings are, of course, very preliminary, but they nonetheless raise policy implications. In particular, efforts should be geared to protecting older adults living in long-term care settings. Policy makers might even consider alternative dwelling options during the epidemic period, such as encouraging residents to live with their families whenever possible.  It appears that additional and more detailed data concerning long-term care facilities may be available in the next few months, allowing us to continue and improve our analysis.

## **Appendix A:**

**Data Sources**

| Description | Source | Comment |
|---|---|---|
| Long-term care beds by country | https://gateway.euro.who.int/en/indicators/hfa_491-5101-number-of-nursing-and-elderly-home-beds/visualizations/#id=19556&tab=table | DK 2011, BE and NL 2012, DE and ES 2013, IE, LU and UK 2014, the rest 2015 |
| Coronavirus death statistics by country | https://www.worldometers.info/coronavirus/ | Data as of May 13, 2020 |
| Demography statistics by countries | https://www.cia.gov/library/publications/the-world-factbook/docs/rankorderguide.html | Year 2020 (Est.) |
| Mobility data | https://www.google.com/covid19/mobility/ | Data through May 9, 2020 |
| Age Groups | https://data.worldbank.org/indicator/SP.POP.65UP.TO.ZS | 2019 Revision |
| Italy Regional long-term care beds | http://dati-anziani.istat.it/index.aspx?lang=en&SubSessionId=83aaf6dc-879c-457e-abe0-ce4781c6f43a | Data as of 2016 |
| Portugal long-term care beds | Lopes, H., Mateus, C, Hernández-Quevedo, C. (2018): Table 2: Ten Years after the Creation of the Portuguese National Network for Long-Term Care in 2006: Achievements and Challenges. Health Policy. | Data as of 2016 |
| Italy regional Coronavirus statistics | https://statistichecoronavirus.it/regioni-coronavirus-italia/toscana/ | Data as of May 15, 2020 |
| Countries density | https://covid.ourworldindata.org/data/owid-covid-data | Year 2020 (Est.) |
| Age group distribution (75+) | https://population.un.org/wpp/Download/Standard/Population/ | Year 2020 (Est.) |
| Hospital Beds by Country | https://data.oecd.org/healtheqt/hospital-beds.htm https://datarepository.wolframcloud.com/resources/OECD-Data-Hospital-Beds-Per-Country | 2016, as reported by the Wolfram Data Repository |